\documentclass[
 aps, pra,
 amsmath,amssymb,
 11pt,
 final,
tightenlines,
 twoside,
 twocolumn,
 nofloats,
nofootinbib,
 superscriptaddress,
showkeys,
showkeywords,
 ]
{revtex4-2}
\usepackage{subfigure}
\usepackage[font={small}]{caption}
\usepackage[T2A]{fontenc}
\usepackage{amsmath}
\usepackage[utf8x]{inputenc}
\usepackage[russian,english]{babel}
\usepackage{graphicx}
\usepackage{dcolumn}
\usepackage{bm}
\usepackage{xcolor}

\input{maik.rty}

\setcitestyle{authoryear,round}
\def\squareforqed{\hbox{\rlap{$\sqcap$}$\sqcup$}}

\def\sq{\ifmmode\squareforqed\else{\unskip\nobreak\hfil
\penalty50\hskip1em\null\nobreak\hfil\squareforqed
\parfillskip=0pt\finalhyphendemerits=0\endgraf}\fi}

\def\utw{\smash{\rlap{\lower5pt\hbox{$\sim$}}}}

\def\udtw{\smash{\rlap{\lower6pt\hbox{$\approx$}}}}

\def\diameter{{\ifmmode\mathchoice
{\ooalign{\hfil\hbox{$\displaystyle/$}\hfil\crcr
{\hbox{$\displaystyle\mathchar"20D$}}}}
{\ooalign{\hfil\hbox{$\textstyle/$}\hfil\crcr
{\hbox{$\textstyle\mathchar"20D$}}}}
{\ooalign{\hfil\hbox{$\scriptstyle/$}\hfil\crcr
{\hbox{$\scriptstyle\mathchar"20D$}}}}
{\ooalign{\hfil\hbox{$\scriptscriptstyle/$}\hfil\crcr
{\hbox{$\scriptscriptstyle\mathchar"20D$}}}}
\else{\ooalign{\hfil/\hfil\crcr\mathhexbox20D}}%
\fi}}

\begin{document}

\selectlanguage{english}



\title{Determination of the Height--Temperature Profile Above a Solar Active Region from Multi-Frequency Radio Observations}

\author{\firstname{T.~I.}~\surname{Kaltman}}
 \affiliation{Special Astrophysical Observatory,
Russian Academy of Sciences, St. Petersburg, 196140 Russia}

\author{\firstname{A.~G.}~\surname{Stupishin}}
 \email{agstup@yandex.ru}
 \affiliation{Special Astrophysical Observatory,
Russian Academy of Sciences, St. Petersburg, 196140 Russia}

\author{\firstname{G.~A.}~\surname{Makoev}}
 \affiliation{Special Astrophysical Observatory,
Russian Academy of Sciences, St. Petersburg, 196140 Russia}

\begin{abstract}
An iterative method is presented for reconstructing the height--temperature profile of the solar atmosphere above a sunspot using multi-frequency spectro-polarimetric microwave observations. It is assumed that the emission is formed predominantly under gyroresonance conditions at harmonics of the electron gyrofrequency, and that the contribution at each frequency is associated with a layer of optical depth of order unity. The frequency--height correspondence is determined from extrapolation of the photospheric magnetic field into the corona. The reconstruction of the profile is reduced to solving an overdetermined system of linear equations with regularization, ensuring noise stability and physical smoothness of the solution. The method is tested on synthetic data for a dipole sunspot model and applied to observations of active region NOAA 11312 obtained with the RATAN-600 radio telescope. The derived temperature profiles are consistent with contemporary models of active regions and reproduce the observed spectra in the 3--18~GHz range with an accuracy of a few percent.
\end{abstract}

\maketitle

\section{INTRODUCTION}

Active regions are areas of concentration of strong and stable magnetic fields in the solar atmosphere \citep{Solanki2003}. The magnetic field determines the plasma stratification and thermodynamic regime from the photosphere to the corona, suppressing convection in the lower layers and directing thermal conduction along magnetic field lines in the transition region and corona \citep{Solanki2006,Borrero2011}. As a result, a characteristic vertical temperature profile is formed above sunspots, substantially different from that of the quiet atmosphere.

In stable (non-flaring) active regions, a quasi-stationary balance between heating, thermal conduction, and radiative losses is established in the transition region and lower corona \citep{Aschwanden2005}. This energy balance determines the height--temperature distribution of the plasma. Therefore, reconstruction of the vertical temperature profile is a key observational task for diagnosing physical conditions and coronal heating mechanisms.

Traditionally, the height--temperature structure of the atmosphere is reconstructed from extreme ultraviolet and X-ray spectra and on the basis of semi-empirical models \citep{Vernazza1981,Maltby1986,Fontenla2009}. However, in the transition region such approaches are complicated by departures from local thermodynamic equilibrium, uncertainties in optical thickness, integration along the line of sight, and the high sensitivity of the energy balance to boundary conditions and heating parameterization \citep{Aschwanden2005}. The nonlinear dependence of thermal conduction and radiative losses on temperature makes the problem ill-conditioned \citep{Spitzer1962}. At coronal heights, the plasma temperature distribution is typically reconstructed from the differential emission measure (DEM), whereas conversion to geometric height requires additional assumptions, and sensitivity to the magnetic field remains limited. This motivates the use of alternative diagnostic methods.

Despite advances in observational techniques, the vertical temperature structure above active regions remains insufficiently constrained. Temperature is not measured directly and is inferred from spectral characteristics of radiation using inversion and model-dependent methods \citep{HannahKontar2012,Cheung2015}. Such methods are simultaneously sensitive to several plasma parameters—density, ionization state, and magnetic field \citep{Zanna2018,Reale2014}—leading to ambiguities in interpretation. Additional complexity arises from the integral nature of observations along the line of sight, especially in the transition region and lower corona, where significant height gradients of temperature and density exist \citep{Judge2010,Slemzin2014,Antonucci2020}.

The microwave range offers several advantages. In the Rayleigh--Jeans approximation, the brightness temperature is directly related to the kinetic temperature of the plasma, enabling quantitative diagnostics \citep{Dulk1985,Gary2004,Lee2007,Shibasaki2011}. Microwave emission is sensitive both to temperature and density (free--free mechanism) and to the magnetic field (gyroresonance mechanism) \citep{Dulk1985,Shibasaki2011,Nindos2020}. In active regions, emission in the several-gigahertz range is formed predominantly at harmonics of the electron gyrofrequency, providing a direct correspondence between observing frequency, magnetic field strength, and formation height \citep{1968SvA....12..245Z,1968SvA....12..464Z,1979SvA....23..316G,1980A&A....82...30A,1996ASSL..204.....Z}. This makes microwave spectra a promising tool for reconstructing the height--temperature structure above sunspots.

In \cite{Stupishin2018}, an iterative method was proposed to match model and observed spectra through successive refinement of the temperature profile. However, the scheme used required a more rigorous mathematical formulation, stability analysis, and a justified regularization procedure.

In the present work, we develop a formalized version of the iterative method for reconstructing the height--temperature profile from microwave spectro-polarimetric observations obtained with the RATAN-600 radio telescope \citep{Parijskij1993,Bogod2011,Bogod2023}. A horizontally homogeneous atmospheric model is considered, in which parameters depend only on height. This allows investigation of the properties of the method under controlled conditions and provides a basis for further model refinement. In contrast to parameterized approaches that prescribe a heating law \citep{Fleishman2021_Heating_Law}, the temperature profile is not imposed a priori.

The method is tested on synthetic data for a dipole model of a sunspot and then applied to observations of active region AR~NOAA~11312. The convergence of the algorithm, its sensitivity to input parameters, and the consistency of the reconstructed profile with contemporary models of the solar atmosphere are analyzed.
\section{METHOD}
\label{sec:method}
\subsection{Physical and Observational Basis}

The method for reconstructing the height--temperature structure of the atmosphere
is based on the properties of microwave emission from active regions
and on the specifics of its formation in magnetized plasma.

In the microwave range, the Rayleigh--Jeans approximation applies,
in which the intensity $I_\nu$ is related to the brightness temperature $T_b$:
\begin{equation}
T_b = \frac{c^2}{k\nu^2}\, I_\nu ,
\end{equation}
where $c$ is the speed of light, $k$ is the Boltzmann constant,
and $\nu$ is the frequency. Under these conditions, $T_b$ directly
characterizes the kinetic temperature of electrons
in the emission formation region.

In the microwave range, the dominant emission mechanism
above sunspots is the gyroresonance (cyclotron) process,
which occurs under the condition
\begin{equation}
\nu \approx s\,\nu_B =
s\,\frac{eB}{2\pi m_e c},
\end{equation}
where $\nu_B$ is the electron gyrofrequency, $B$ is the magnetic
field strength, $e$ and $m_e$ are the electron charge and mass,
and $s$ is the harmonic number.
In active regions, the main contribution typically arises
from the second and third (more rarely the fourth) harmonics.

Since the magnetic field decreases with height,
different frequencies are formed at different atmospheric levels,
which establishes a correspondence between observing frequency
and formation height and makes the microwave spectrum sensitive
to the vertical structure of temperature and magnetic field.

Implementation of the method requires multi-frequency
spectro-polarimetric observations.
Recording right- and left-circular polarizations
makes it possible to account for differences in the formation conditions
of emission in the ordinary and extraordinary modes,
thereby increasing the diagnostic capability.

In this work, observations from the RATAN-600 radio telescope
in the $3$--$18$~GHz range are used. Despite the one-dimensional and frequency-dependent nature of the spatial resolution, the broad frequency coverage, spectral sampling, registration of right- and left-circular polarizations, and the high sensitivity of the instrument provide sufficient diagnostic capability to address the reconstruction of the height--temperature structure within the adopted approximation.

The proposed approach is not limited to RATAN-600 observations
and can in the future be generalized to two-dimensional data
from interferometric instruments
(SRH, EOVSA, etc.).

\subsection{Atmospheric Model}

For testing the method under controlled conditions, a horizontally homogeneous plane-parallel atmospheric model is introduced, in which plasma parameters depend only on height $h$.
In the layers where microwave emission is formed,
local thermodynamic equilibrium is assumed,
so that the brightness temperature characterizes
the electron temperature of the plasma.

The atmosphere is discretized into $M$ horizontal layers
with heights $h_i$ (measured from the photospheric level),
temperature $T_i$, and electron density $N_i$.
The height step is specified as an input parameter:
it is reduced in regions of steep temperature gradients
and increased in quasi-isothermal layers.

The height distribution of electron density above some height $h_0$ 
is described by a barometric law
(\cite{Aschwanden2005}, \S~3.1):
\begin{equation}
N(h) = N(h_0)\,\frac{T(h_0)}{T(h)}
\exp\!\left(-\frac{h-h_0}{\lambda T(h)}\right),
\label{barometric}
\end{equation}
where $\lambda = 4.7 \times 10^3 \,\mathrm{cm\,K^{-1}}$;
the formula is valid for heights much smaller than the solar radius.

For a layer between $h_1$ and $h_2$, the contribution to the flux
at frequency $\nu$ in polarization $p$
( RCP or LCP ) is determined by the optical depth
\begin{equation}
\tau_{\nu,p} = \int_{h_1}^{h_2}
\kappa_{\nu,p}(h)\,dh,
\end{equation}
where $\kappa_{\nu,p}$ is the absorption coefficient,
which depends on the plasma parameters and frequency.

The dominant contribution to the emission at a given frequency
comes from the layer where $\tau_{\nu,p} \sim 1$,
which establishes the correspondence between frequency
and the characteristic formation height.

The model flux is calculated on the basis of
synthetic radio maps representing
the spatial distribution of brightness.
These maps are convolved with the RATAN-600 beam pattern,
yielding one-dimensional scans,
which are compared with the observed spectra
at the position of maximum intensity.

\subsection{Iterative Algorithm}
\label{sec:iterations}


In this section, we describe the algorithm for solving the inverse problem of reconstructing the height--temperature profile from observed microwave spectra. The algorithm is based on iterative refinement of the temperature distribution in order to minimize the residual between the model and observed emission.

The reconstruction problem is formulated as the determination of a height--temperature profile that provides the best agreement between the model spectrum $F^{\mathrm{model}}$ and the observed spectrum $F^{\mathrm{obs}}$.

It is important to note that the proposed algorithm requires flux measurements in polarization, since in different propagation modes the optical depth $\tau = 1$ may be reached at different harmonics of the gyrofrequency and, consequently, at different heights.

For a given temperature profile, we compute a radio map of the region by defining a coordinate grid ${X, Y}$. Since the dominant contribution to the flux comes from the layer where $\tau \approx 1$, the model flux $F_{\nu,p,i}$ at frequency $\nu$ for each polarization $p$, originating from layer $h_i$, can be represented as a pixel-by-pixel sum of contributions from those pixels of the radio map where the optical depth reaches $\tau \approx 1$:

\begin{figure}
\includegraphics[width=1\linewidth]{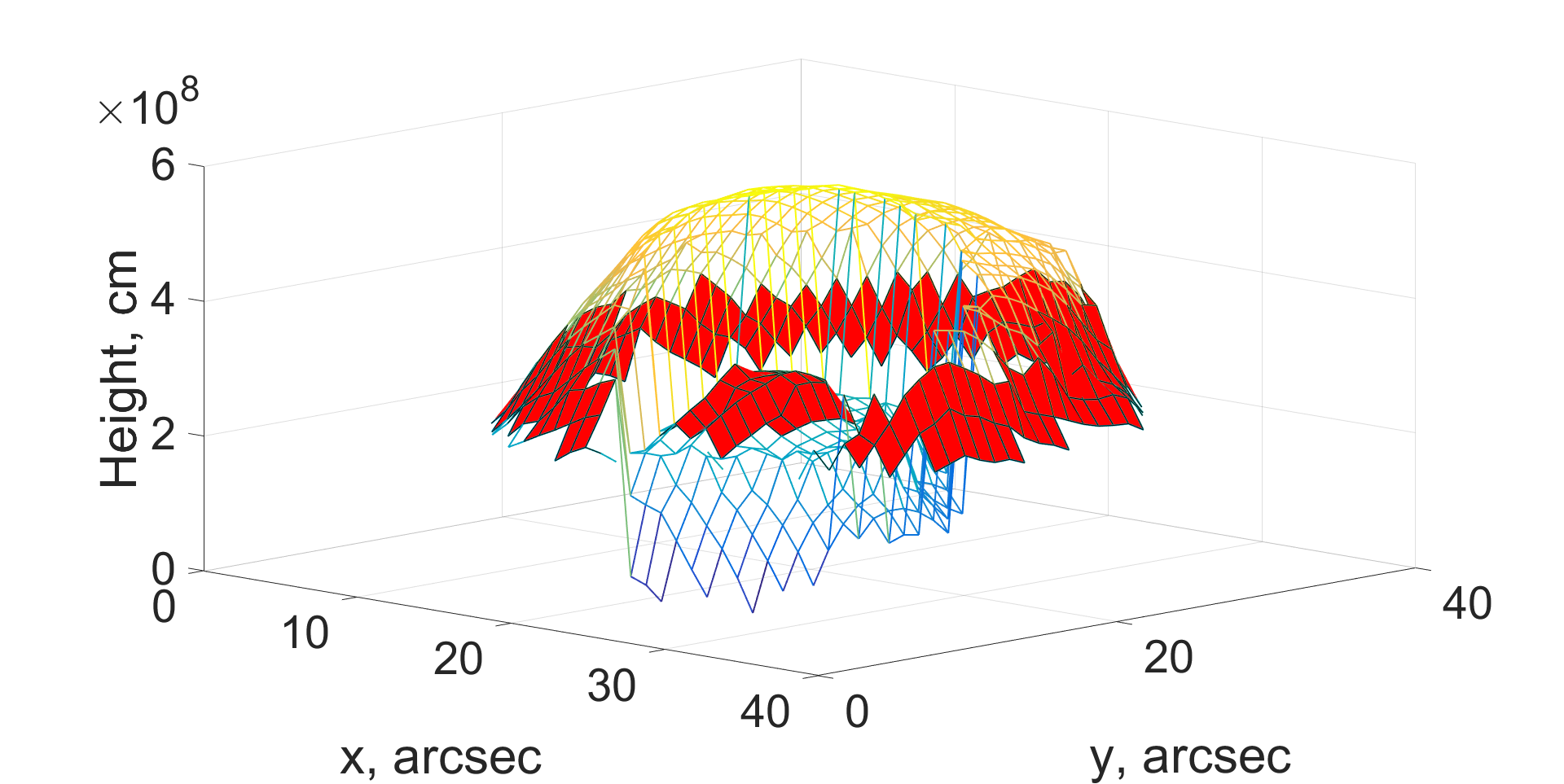}
\includegraphics[width=1\linewidth]{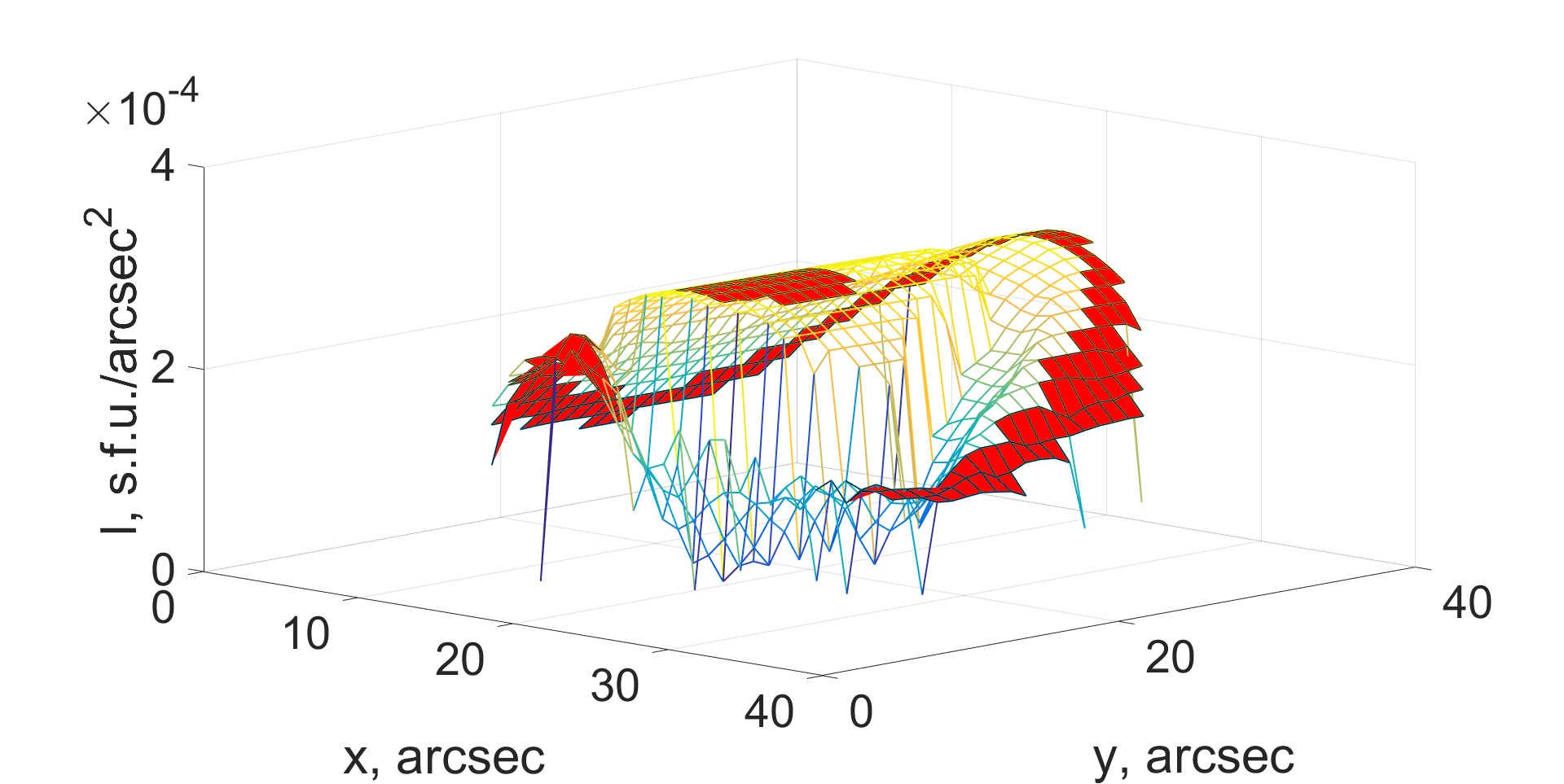}
\caption{Top panel: surfaces of the gyrolevels corresponding to the second and third harmonics; 
pixels of layer $h_i$ where $\tau \approx 1$ are highlighted in red. 
Bottom panel: intensity surface. 
The summation in Equation~(\ref{eq:summa_tau_1}) is performed over the pixels marked in red.
}
\label{ris:pixels_tau_1}
\end{figure}

\begin{equation}
F_{\nu,p,i} = \sum_{x,y: \, \tau \approx 1} \, F_{\nu,p,i}(x, y),
\label{eq:summa_tau_1}
\end{equation}

where $x, y$ are the coordinates on the grid ${X, Y}$ (see Fig.~\ref{ris:pixels_tau_1}). The total model flux $F_{\nu,p}$ is then obtained by summing over the height layers:

\begin{equation}
F_{\nu,p}^{\mathrm{model}} = \sum_{i=1}^M \, F_{\nu,p,i} + F_{\nu,p}^{\mathrm{thin}},
\label{eq:summa_by_layers}
\end{equation}

\begin{equation}
F_{\nu,p}^{\mathrm{thin}} = \sum_{x,y: \, \tau < 1} \, F_{\nu,p}^{} .
\end{equation}

Here, $F_{\nu,p}^{\mathrm{thin}}$ represents the contribution from the optically thin part of the atmosphere, which cannot be uniquely associated with a specific height layer.

Our goal is to achieve the closest possible agreement between the computed ($F_{\nu,p}^{\mathrm{model}}$) and observed ($F_{\nu,p}^{\mathrm{obs}}$) fluxes, i.e., to satisfy
\begin{equation}
F_{\nu,p}^{\mathrm{model}} = F_{\nu,p}^{\mathrm{obs}}, \qquad \forall \nu, p
\end{equation}
or, taking into account expression (\ref{eq:summa_by_layers}),
\begin{equation}
\sum_{i=1}^M \, F_{\nu,p,i} = F_{\nu,p}^{\mathrm{obs}} - F_{\nu,p}^{\mathrm{thin}}. \qquad \forall \nu, p
\label{eq:equal_fluxes}
\end{equation}

Since the initial height--temperature profile is specified arbitrarily, the corresponding model flux is, in general, not consistent with the observed spectrum: some height layers will contribute less than the actual physical layer, while others will contribute more. We therefore need to develop a mechanism that allows us to adjust the contributions from different height layers by correcting the layer temperatures.

Let us introduce a set of coefficients $\alpha_i$ ($i=1 \dots M$) to correct the temperature at the $k$-th iteration:
\begin{equation}
T_{\nu,p,i}^{(k+1)} = \alpha_i\, T_{\nu,p,i}^{(k)}.
\end{equation}

Under local thermodynamic equilibrium (LTE), the flux scales as $F \sim T$, so we can write the previous relation as
\begin{equation}
F_{\nu,p,i}^{(k+1)} = \alpha_i\, F_{\nu,p,i}^{(k)}.
\label{eq:corr_coefs}
\end{equation}

Equations~(\ref{eq:equal_fluxes}) and (\ref{eq:corr_coefs}) lead to a system of linear equations for determining the correction coefficients:
\begin{equation}
\mathbf{F}\,\vec{\alpha} = \vec{f},
\label{eq:SLEBase}
\end{equation}

The matrix $\mathbf{F}$ has the form
\begin{equation}
\mathbf{F} =
\left(
\begin{matrix}
R^{(1)}_1 & R^{(2)}_1 & \hdots & R^{(M)}_1 \\
R^{(1)}_2 & R^{(2)}_2 & \hdots & R^{(M)}_2 \\
\hdots & \hdots & \ddots & \hdots \\
R^{(1)}_{N_R} & R^{(2)}_{N_R} & \hdots & R^{(M)}_{N_R} \\
L^{(1)}_1 & L^{(2)}_1 & \hdots & L^{(M)}_1 \\
L^{(1)}_2 & L^{(2)}_2 & \hdots & L^{(M)}_2 \\
\hdots & \hdots & \ddots & \hdots \\
L^{(1)}_{N_L} & L^{(2)}_{N_L} & \hdots & L^{(M)}_{N_L}
\end{matrix}
\right),
\end{equation}
where $R^{(j)}_k$ and $L^{(j)}_k$ are the contributions of the $i$-th layer to the model flux at the $j$-th frequency in right- and left-circular polarizations, respectively.

The vector of correction factors is
\begin{equation}
\vec{\alpha} =
\left(
\begin{matrix}
\alpha_1 \\
\alpha_2 \\
\vdots \\
\alpha_M
\end{matrix}
\right),
\end{equation}

and the vector on the right-hand side  of the system is defined as 

\begin{equation}
\vec{f} =
\left(
\begin{matrix}
R^{\mathrm{obs}}_{1} - R^{\mathrm{thin}}_{1} \\
R^{\mathrm{obs}}_{2} - R^{\mathrm{thin}}_{2} \\
\vdots \\
R^{\mathrm{obs}}_{N_R} - R^{\mathrm{thin}}_{N_R} \\
L^{\mathrm{obs}}_1 - L^{\mathrm{thin}}_1 \\
L^{\mathrm{obs}}_2 - L^{\mathrm{thin}}_2 \\
\vdots \\
L^{\mathrm{obs}}_{N_L} - L^{\mathrm{thin}}_{N_L}
\end{matrix}
\right).
\end{equation}

${N_R}, {N_L}$ denote the numbers of observed (and modeled) frequencies in right and left circular polarization, respectively.

To stabilize the solution and ensure its physical plausibility, regularization is introduced in the form of a smoothness condition on the temperature profile, requiring suppression of sharp temperature variations between adjacent layers:

\begin{equation}
\alpha_i T_i = \alpha_{i+1} T_{i+1} \qquad \forall\, i .
\end{equation}

This condition is added to the system of linear equations (\ref{eq:SLEBase}). In its final form, the system becomes

\begin{equation}
\Theta \vec \alpha = \vec \phi, 
\end{equation}
where
\begin{equation}
\Theta = \left( 
\begin{matrix}
\textbf{F} \\
\textbf{T} \\
\end{matrix}
\right), \quad \vec \phi = \left( 
\begin{matrix}
\vec f \\
\vec 0 \\
\end{matrix}
\right),
\end{equation}

\textbf{T} is a bidiagonal matrix of size $(M-1) \times M$:

\begin{equation} \label{SLE_2diagT}
\textbf{T} =
\left( 
\begin{matrix}
T_1    & -T_2   &  0     & \hdots & 0         & 0      \\
0      &  T_2   & -T_3   & \hdots & 0         & 0      \\
\hdots & \hdots & \hdots & \ddots & \hdots    & \hdots \\
0      &  0     &  0     & \hdots & T_{M-1}   & -T_M   \\
\end{matrix}
\right).
\end{equation}

This augmented system is overdetermined (since it has $M$ columns and $N_R+N_L+M-1$ rows), and therefore the solution is obtained using the least-squares method. The solution has the form

\begin{equation}
\vec \alpha = (\Theta' \textbf{W} \Theta)^{-1} \Theta' \textbf{W} \vec \phi
\label{eq:LMS}
\end{equation}

where $\mathbf{W}$ is a weighting matrix. In the simplest case, it is the identity matrix. However, if a priori estimates of the relative importance of individual equations are available, the weights of selected equations can be reduced (for example, lower weights may be assigned to noisier frequencies, etc.).

The iterative procedure for reconstructing the height--temperature profile is organized as follows:
\begin{enumerate}
  \label{list:iterations}
  \item An initial approximation to the height--temperature profile $\{T_i^{(0)}\}$ is specified.
  
  \item At the $k$-th iteration, using the current temperature profile $\{T_i^{(k)}\}$, the model microwave flux $F_{\nu, p}^{\mathrm{model}}$ and its components $F_{\nu,p,i}^{\mathrm{model}}$, $F_{\nu,p}^{\mathrm{thin}}$ are calculated at all considered frequencies and in both circular polarizations (Equations~(\ref{eq:summa_tau_1}), (\ref{eq:summa_by_layers})).

  \item By solving Equation~(\ref{eq:LMS}), the vector of correction factors $\{\alpha_i^{(k)}\}$ is determined.
  
  \item The temperature profile for iteration $k+1$ is updated according to
  \begin{equation}
  T_i^{(k+1)} = \alpha_i^{(k)}\, T_i^{(k)} .
  \end{equation}
  
  \item The iterations are continued until stabilization of the residual between the model and observed spectra is achieved.
The stopping criterion is that the change in  the residual falls below a prescribed threshold over several consecutive steps.

\end{enumerate}

Convergence of the algorithm is typically achieved within $10$--$30$ iterations.

\section{TESTING WITH A DIPOLE MODEL}
\subsection{Model Description}

For the primary verification of the iterative method, a simplified sunspot model with a prescribed magnetic configuration and height--temperature atmospheric profile is used. The use of synthetic model data makes it possible to test the correctness of the algorithm under controlled conditions, where the true temperature distribution is known.

Numerical modeling of the microwave emission was performed using the gyroresonance emission calculation package developed in \citep{Stupishin2018}.

The magnetic field in the model is specified as a vertical magnetic dipole located at the center of the solar disk and buried below the photosphere at a depth
$h_d = 16\,\mathrm{Mm}$.

This approximation provides a realistic decrease of the magnetic field strength with height and reproduces the characteristic magnetic configuration of isolated sunspots \citep{Lites1990,Zlotnik1996,Brosius1989}.

The maximum magnetic field strength at the base of the sunspot, $B_{\max}$, is considered in the range $2000$--$3000$~G.  In addition, the spatial resolution of the computational grid of the radio map is varied in order to assess the influence of numerical parameters on the modeling results.

The height--temperature profile of the atmosphere is specified in the form of three characteristic regions:
a cold, dense chromosphere up to a height of $h \approx 1.5\,\mathrm{Mm}$ with temperature
$T \simeq 10^{4}$~K;
a transition region at heights $1.5$--$1.8\,\mathrm{Mm}$ with a sharp temperature increase toward coronal values
$T \simeq 2.5 \times 10^{6}$~K;
and a hot, rarefied corona at $h > 1.8\,\mathrm{Mm}$.

The height distribution of the electron density $N(h)$ is described by a barometric law (\ref{barometric}), based on a prescribed density at the base of the corona and a characteristic density scale height.

\subsection{Generation of Synthetic Data}

\begin{figure}
\includegraphics[width=0.8\linewidth]{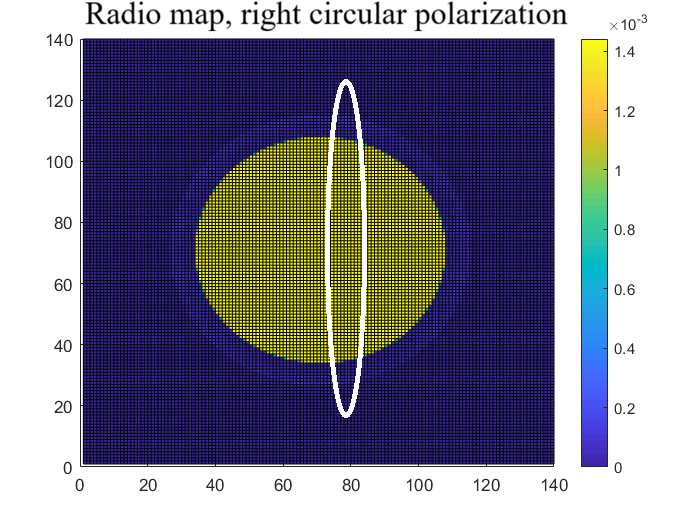}
\includegraphics[width=0.8\linewidth]{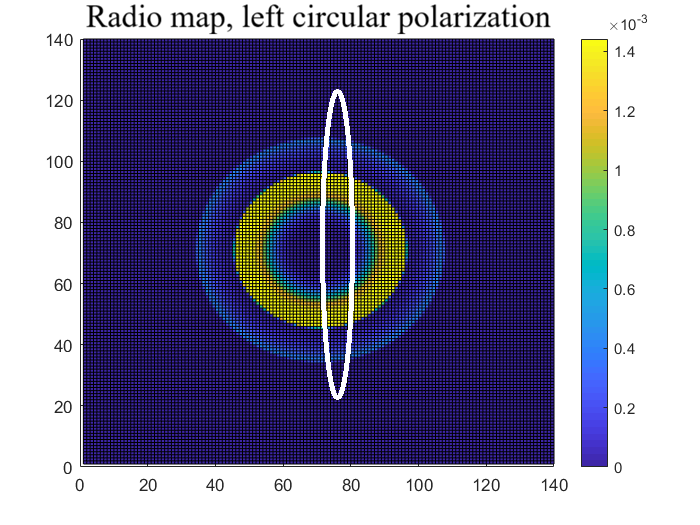}
\includegraphics[width=0.7\linewidth]{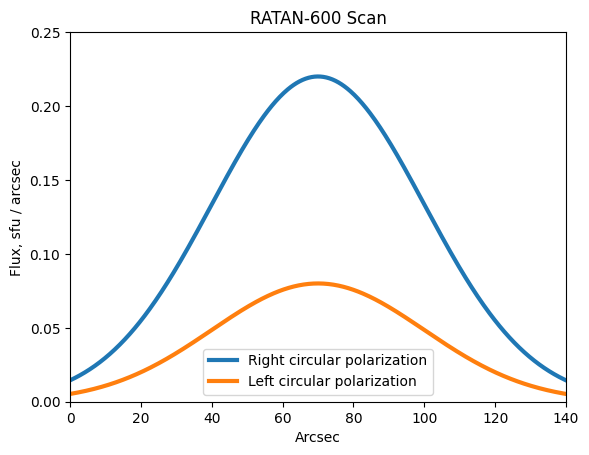}
\caption{Top: calculated radio maps of the active region for the dipole magnetic-field model
($B_{\max}=3000$~G, $f=8$~GHz) in right- and left-circular polarizations.
Bottom: one-dimensional scans obtained by convolving the radio maps with the RATAN-600 beam pattern.}
\label{ris:Radomap_scan}
\end{figure}

Based on the prescribed magnetic configuration and height--temperature profile, numerical modeling of the microwave emission of the active region is performed. For each frequency and each circular polarization, radio maps of the brightness temperature are calculated, taking into account the gyroresonance emission mechanism. The radio maps are then convolved with the RATAN-600 beam pattern, yielding one-dimensional scans analogous to real observations (Fig.~\ref{ris:Radomap_scan}).

From these scans, the maximum flux values at each frequency are extracted; their collection forms the model microwave spectrum in right- and left-circular polarizations. This spectrum is used as a test input for the iterative reconstruction algorithm of the height--temperature profile.

The testing is carried out for various initial approximations, including substantially shifted profiles (for example, with an artificially elevated transition-region height), allowing assessment of the stability and convergence of the method.

\subsection{Test Results}

\begin{figure*}
\includegraphics[width=1\linewidth]{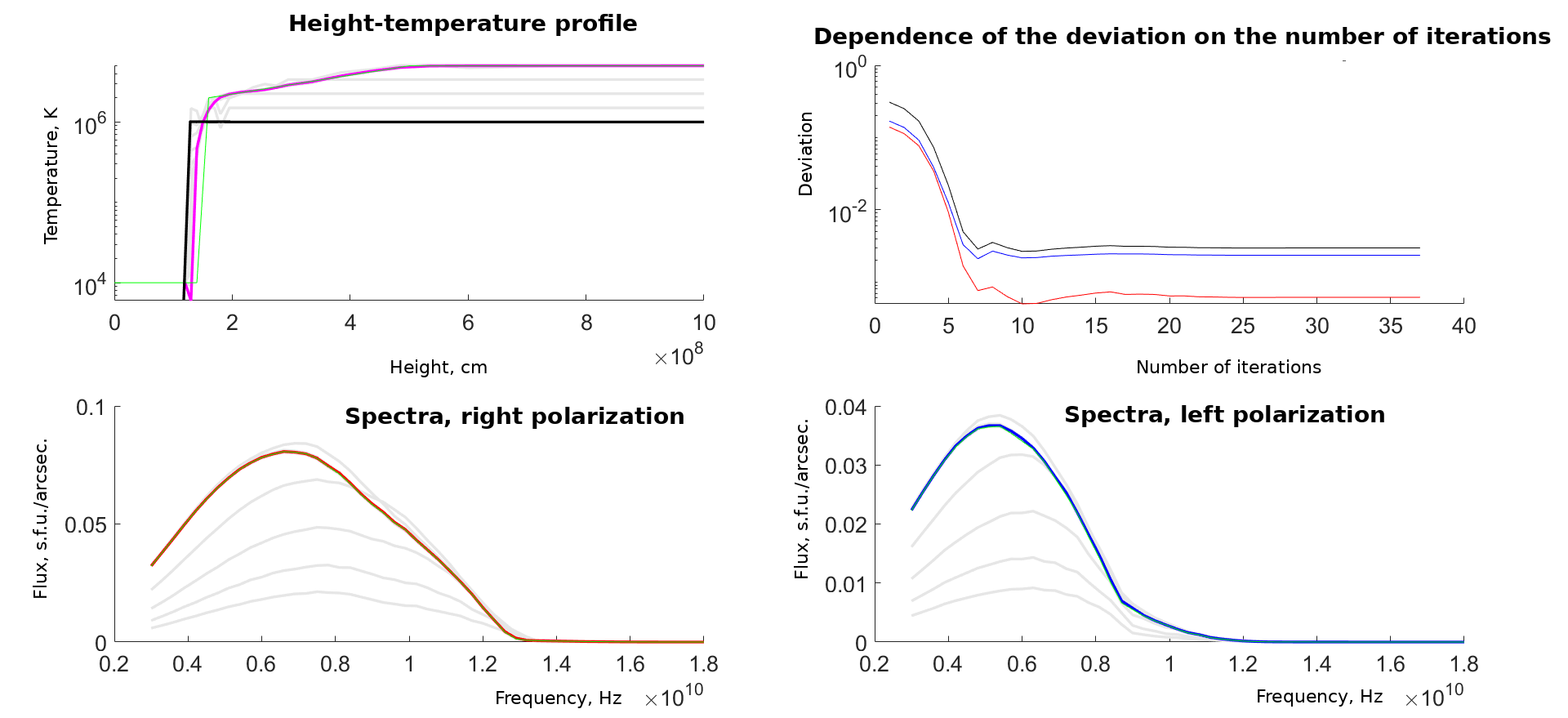}
\includegraphics[width=1\linewidth]{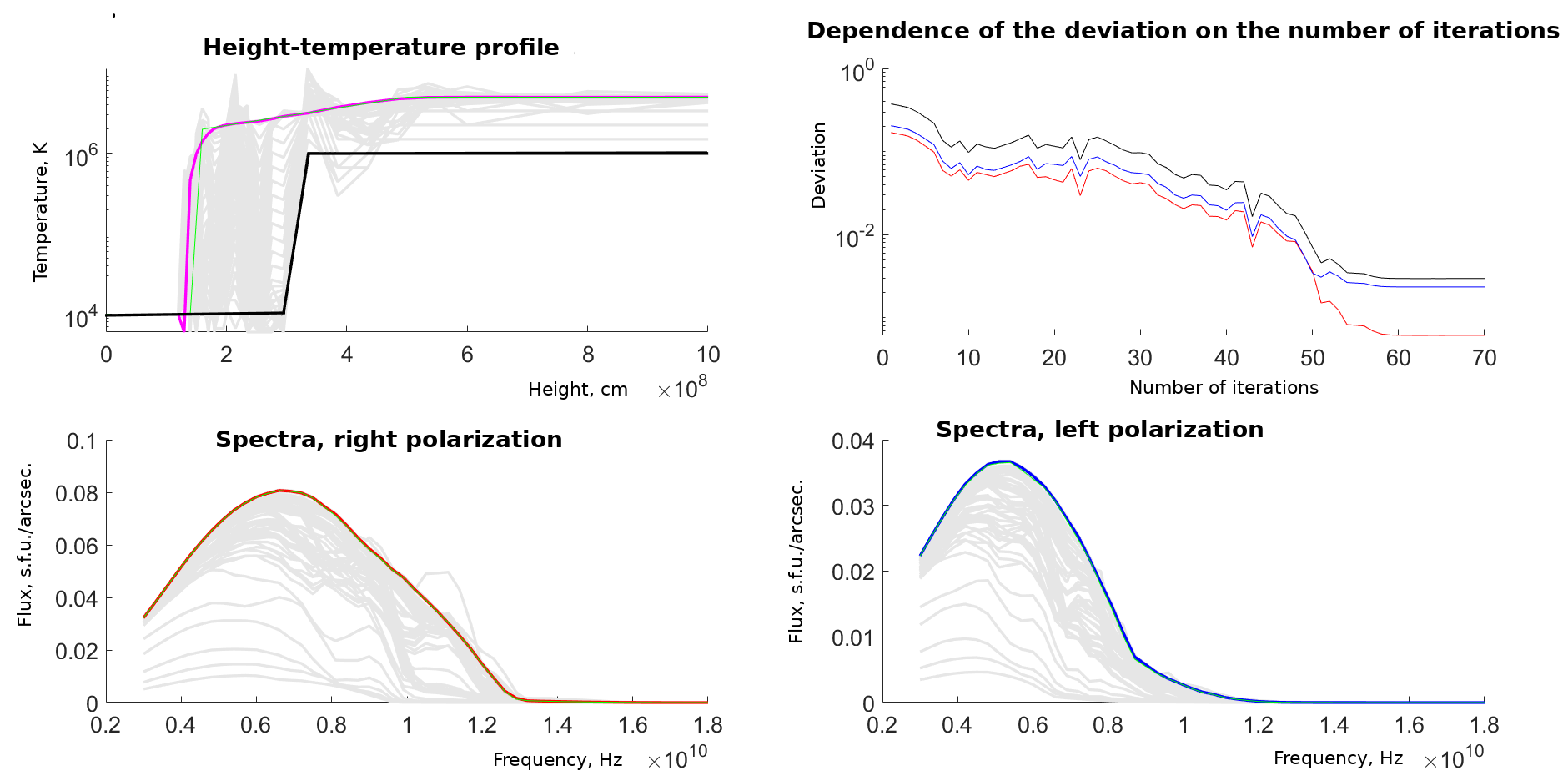}
\caption{Results of testing the iterative algorithm for reconstructing the height--temperature profile
for a dipole model of an active region.
The upper panels show the result for an initial temperature profile close to the model profile,
while the lower panels correspond to an initial profile with an artificially elevated transition-region height.
In each case, the height--temperature profile, the dependence of the residual on iteration number,
and the microwave spectra in right- and left-circular polarizations are shown.}
\label{fig:Test}
\end{figure*}

The results of the algorithm testing are presented in Fig.~\ref{fig:Test}. The upper set of panels corresponds to the case of an initial profile close to the model profile, while the lower set corresponds to a substantially shifted initial approximation with an increased transition-region height.

In the panels showing the height--temperature profiles, the green curve represents the true temperature distribution, the black curve shows the initial approximation, and the magenta curve corresponds to the reconstructed profile after completion of the iterations. The gray curves represent intermediate iterations and illustrate the convergence trajectory of the algorithm.

The residual plots demonstrate a decrease in error with increasing iteration number. The red and blue curves correspond to the residuals in right- and left-circular polarizations, respectively, while the black curve represents their combined residual. In both tests, a stable reduction of the residual is observed; for the case of an initial profile close to the model, convergence is achieved in a smaller number of iterations, whereas for a substantially shifted initial approximation more iterations are required. Nevertheless, the final residual level in both cases decreases to $\sim 0.3\,\%$.

In the spectral panels, the green curves correspond to the model (true) spectra, while the red and blue curves show the reconstructed spectra in right- and left-circular polarizations, respectively. The gray spectra represent intermediate iterations and demonstrate the gradual convergence toward the model distribution.

Thus, the algorithm successfully reproduces the prescribed height--temperature profile even for a substantially distorted initial approximation, while maintaining stability and controlled convergence.

\subsection{Sensitivity of the Method to Noise in the Input Spectra}
\label{sec:noise_sensitivity}

\begin{figure*}[t]
  \centering
  \begin{tabular}{cc}
    \includegraphics[width=0.48\textwidth]{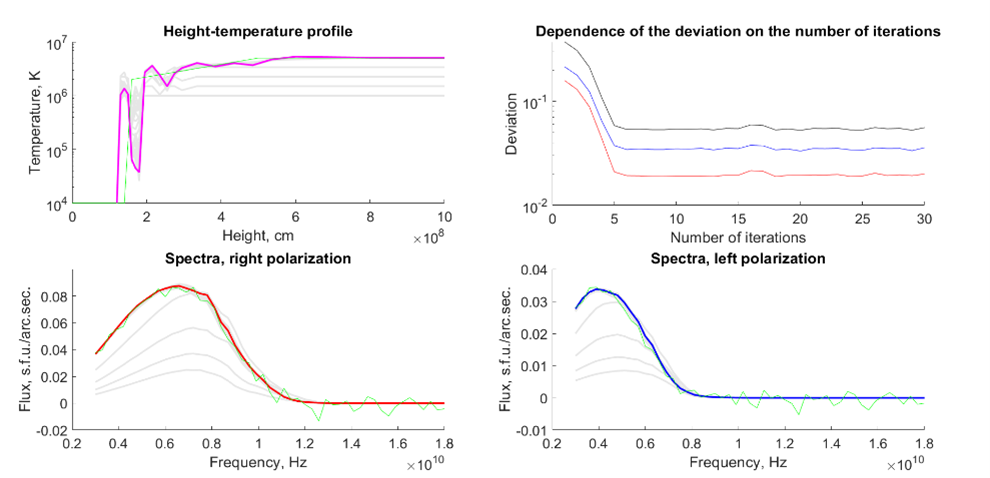} &
    \includegraphics[width=0.48\textwidth]{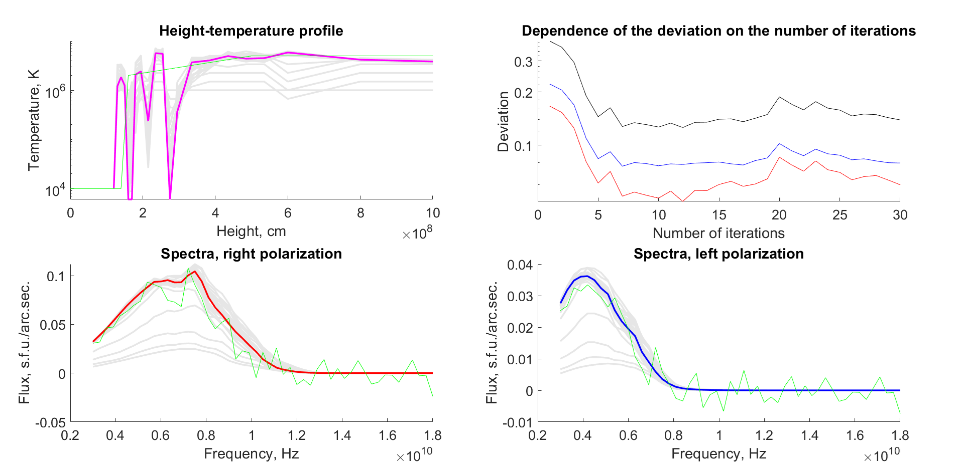}
  \end{tabular}
  \caption{
  Testing the sensitivity of the iterative method to additive noise
  in the input spectra.
  The reference and reconstructed height--temperature profiles are shown,
  together with the quality of spectral reproduction in right- and left-circular
  polarizations.
  (a) additive noise with a level of $5\%$;
  (b) additive noise with a level of $10\%$.
  }
  \label{fig:noise_add}
\end{figure*}

\begin{figure*}[t]
  \centering
  \begin{tabular}{cc}
    \includegraphics[width=0.48\textwidth]{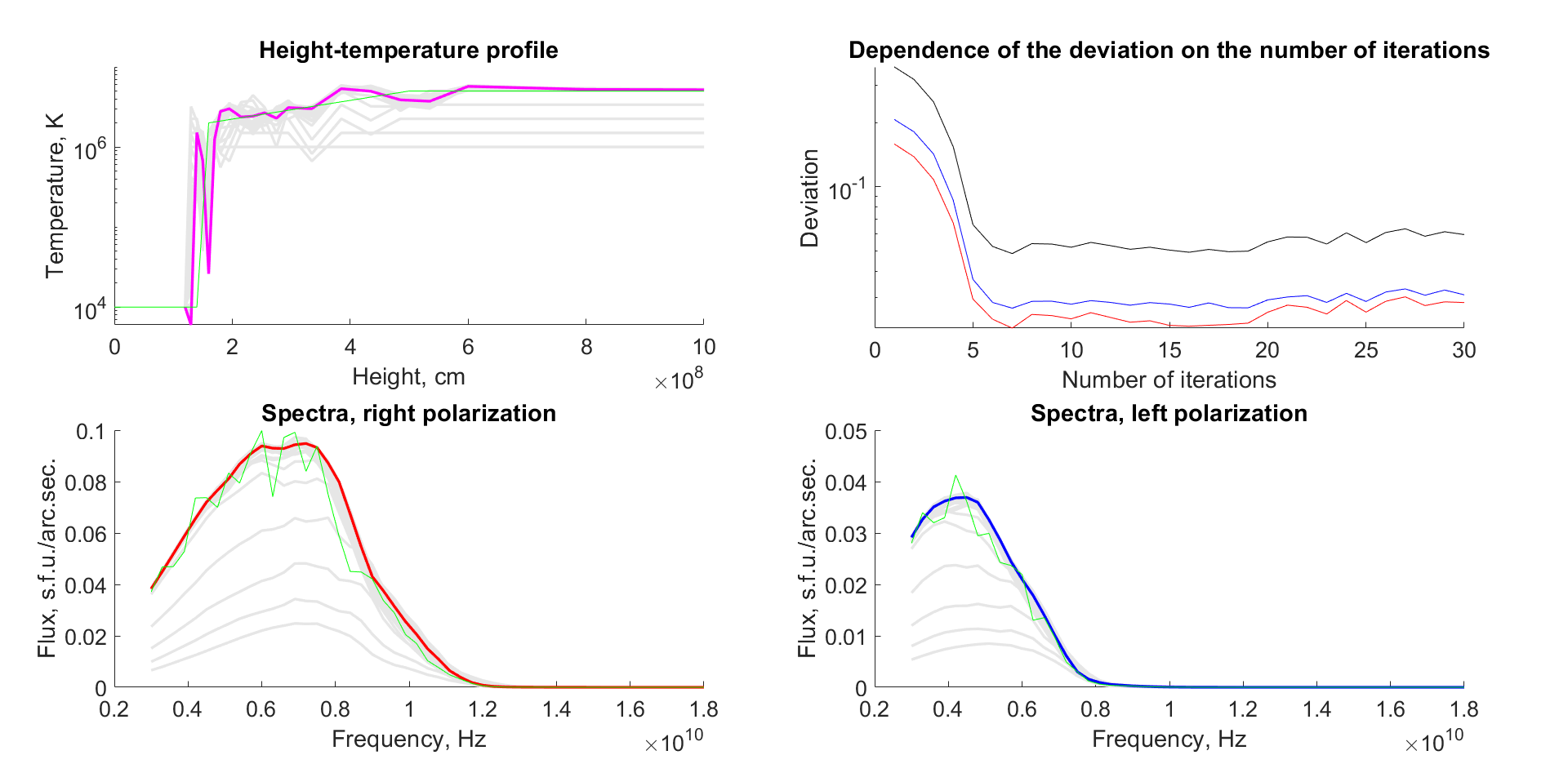} &
    \includegraphics[width=0.48\textwidth]{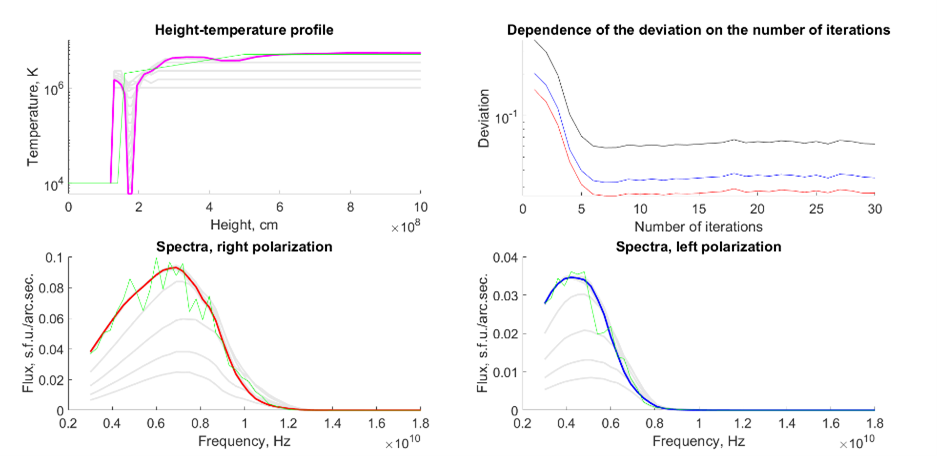}
  \end{tabular}
  \caption{
  Testing the robustness of the iterative method to multiplicative noise
  and the effect of enhanced regularization.
  The reference and reconstructed height--temperature profiles are shown,
  together with the reproduced spectra in right- and left-circular polarizations.
  (a) multiplicative noise with a level of $10\%$;
  (b) the same noise level with the weights of the regularization
  (temperature) equations increased by a factor of 100.
  }
  \label{fig:noise_mult_reg}
\end{figure*}

\label{sec:noise}

To assess the robustness of the method to distortions in the input spectra,
artificial noise of a prescribed type and level was added to the reference model spectrum,
after which the height--temperature profile was reconstructed using the same algorithm.

Two types of errors were considered: additive
(receiver noise and background fluctuations)
and multiplicative (calibration errors and gain instability).
Additive noise was introduced as
\[
S_i' = S_i + \xi_i\, \sigma\, S_{\max},
\]
and multiplicative noise as
\[
S_i' = S_i (1 + \xi_i\, \sigma),
\]
where $\sigma$ is the relative noise level
and $\xi_i \in [-1,1]$ is a random variable.

For low noise levels ($\sigma=1\%$, not shown),
the reconstructed profile practically coincides with the reference one;
differences are limited to small variations in the transition region.
For additive noise with $\sigma=5\%$,
distortions in the position and steepness of the transition region become noticeable,
while the coronal part of the profile remains stable.
At $\sigma=10\%$, the transition region becomes the main source of uncertainty;
the noise contribution becomes the dominant factor limiting
the reconstruction accuracy in this layer
(Fig.~\ref{fig:noise_add}).

In the case of multiplicative noise, the reconstruction proves to be more robust.
Even at $\sigma=10\%$, the overall shape of the profile is preserved,
and distortions of the transition region are less pronounced
(Fig.~\ref{fig:noise_mult_reg}a).
This is because multiplicative errors preserve the relative spectral shape,
whereas additive noise introduces a frequency-independent amplitude offset
that more strongly distorts the spectral structure of the signal.

At high noise levels, stability of the reconstruction
is achieved by strengthening the regularization.
Increasing the weights of the temperature equations leads to smoothing
of the reconstructed profile and suppression of nonphysical oscillations,
primarily in the transition region, but is accompanied by an increase
in the spectral residual, reflecting a trade-off between solution smoothness
and fidelity to the noisy data
(Fig.~\ref{fig:noise_mult_reg}b).
The example shown in the figure corresponds to multiplicative noise;
however, an analogous result was obtained for additive noise
at the same level $\sigma = 10\%$ when the weights of the regularization equations
were increased by a factor of 100.

The main conclusions of the testing are that the method remains operational
under moderate noise levels in the input spectra;
the transition region is the most sensitive part of the reconstruction;
multiplicative noise affects the result less strongly than additive noise
at comparable levels; and strengthening regularization allows stabilization
of the solution under strong noise at the cost of increased residual.

\section{APPLICATION OF THE METHOD TO OBSERVATIONS OF ACTIVE REGION NOAA~11312}

\begin{figure*}
	\includegraphics[width=0.99\textwidth]{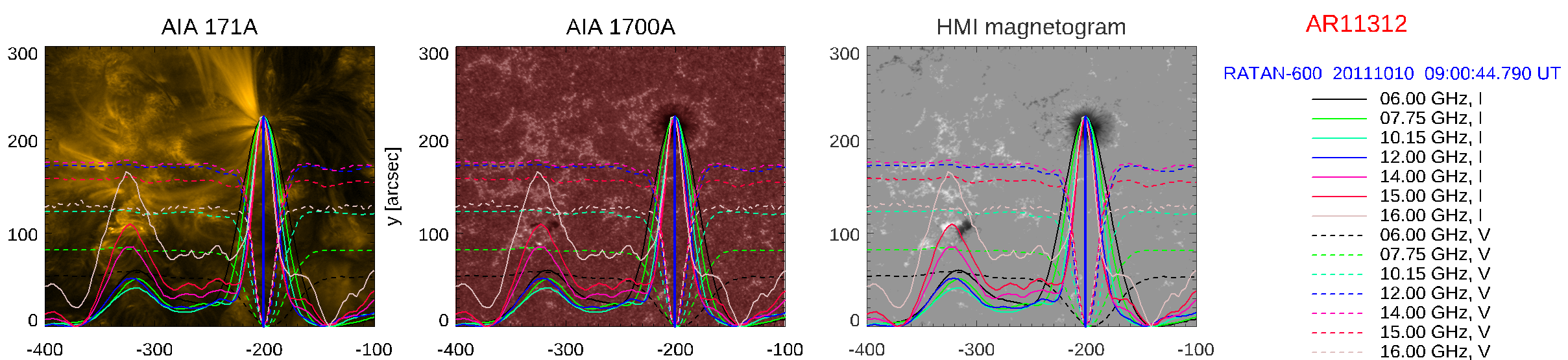}
\caption{
Active region AR11312 (10 October 2011). From left to right: SDO/AIA 171\,\AA, AIA 1700\,\AA\ images and the SDO/HMI magnetogram. Colored curves show one-dimensional RATAN-600 scans at selected frequencies indicated in the legend, spatially co-aligned with the images: solid lines correspond to Stokes~I, dashed lines to Stokes~V. The profiles are scaled in amplitude to the panel height; polarization profiles (Stokes~V) are additionally shifted vertically for clarity.
}
	\label{ar11312_sdo_scans}
\end{figure*}

\subsection{Observational Data}

To apply the developed method to real solar observations,
active region NOAA~11312 was selected
(Fig.~\ref{ar11312_sdo_scans}).
This region represents a relatively isolated single sunspot
with a simple, nearly axisymmetric structure, which previously motivated
its selection as a target for analysis in
\cite{Stupishin2018,Alissandrakis2019}.
Owing to its morphological simplicity and the absence of a pronounced
complex bipolar structure, NOAA~11312 provides a convenient test object
for applying a horizontally homogeneous model in which atmospheric parameters
are assumed to depend only on height and are averaged over the sunspot cross-section.
This enables consistent comparison of different diagnostic approaches
and allows the region to be used as a reference case for validating the algorithm.

In SDO/AIA 1700\,\AA\ images, the sunspot (area $\sim 210$~msh)
with a symmetric penumbra is clearly visible, while in the AIA 171\,\AA\ channel
a system of coronal loops is observed above the spot,
localized predominantly over its central part.
The SDO/HMI magnetogram shows the dominance of a single polarity within the sunspot
and weak magnetic fields outside it,
consistent with the character of circular polarization in the microwave scans.

Radio observations were performed on 10 October 2011 with the RATAN-600 radio telescope
in the frequency range $3$--$18$~GHz, with measurements of the Stokes parameters $I$ and $V$.
In solar observations, the spatial resolution of RATAN-600
is one-dimensional and determined by the beam size,
which depends on wavelength. The beam dimensions are given by
$Q_{\mathrm{hor}}(\mathrm{arcsec}) = 0.85\,\lambda(\mathrm{mm})$
and
$Q_{\mathrm{vert}}(\mathrm{arcmin}) = 0.75\,\lambda(\mathrm{mm})$,
where $Q_{\mathrm{hor}}$ is the horizontal (scan-direction) beam size,
$Q_{\mathrm{vert}}$ is the vertical beam size, and $\lambda$ is the observing wavelength in mm.
In the considered frequency range, this corresponds
to a horizontal resolution from $\sim 15''$ (18~GHz)
to $\sim 90''$ (3~GHz).
The one-dimensional scans, co-aligned with the ultraviolet images
and magnetogram, are also shown in Fig.~\ref{ar11312_sdo_scans}.
Primary data processing included standard calibration
and subtraction of the quiet-Sun background emission.

For modeling the microwave emission, a cutout
of the active region from the full-disk HMI magnetogram was used.
The photospheric magnetic field was extrapolated under the assumption
of a nonlinear force-free field (NLFFF; \citealt{2023ApJS..267....6N}),
yielding a three-dimensional magnetic-field configuration
above the sunspot, which was then used in the gyroresonance emission calculations.

\subsection{Construction of the Input Spectrum and Modeling Parameters}


One of the scans, namely that obtained at 15~GHz, was used to determine the position of the maximum radio flux of the active region. The flux values at this position for all observing frequencies and for both circular polarizations (RCP and LCP) formed the observed microwave spectrum $F_{\nu,p}^{\mathrm{obs}}$, which was used as input to the iterative algorithm. Thus, the input vector $F_{\nu,p}^{\mathrm{obs}}$ consisted of the set of amplitudes at the scan maximum.

Based on the reconstructed magnetic field and a prescribed initial height--temperature profile, model one-dimensional scans were calculated, and the iterative procedure described in Section~\ref{sec:iterations} was applied.

One of the key model parameters was the product of electron density and temperature at the base of the corona, $NT$, which determines the electron pressure and affects the height distribution of density. The value of $NT$ was varied within physically reasonable limits in order to achieve the best agreement between the model and observed spectra. For each selected value of $NT$, the iterative algorithm was run independently, and the solution corresponding to the minimum residual was chosen. In the present case, the best agreement was achieved for
\[
NT = 1 \times 10^{16}\ \mathrm{K\,cm^{-3}}.
\]

\subsection{Reconstruction Results for Active Region NOAA~11312}

Application of the iterative method to observations of active region
NOAA~11312 made it possible to reconstruct the height--temperature profile
of the atmosphere above the sunspot within the framework of the homogeneous model.

\begin{figure}
\center{\includegraphics[width=1\linewidth]{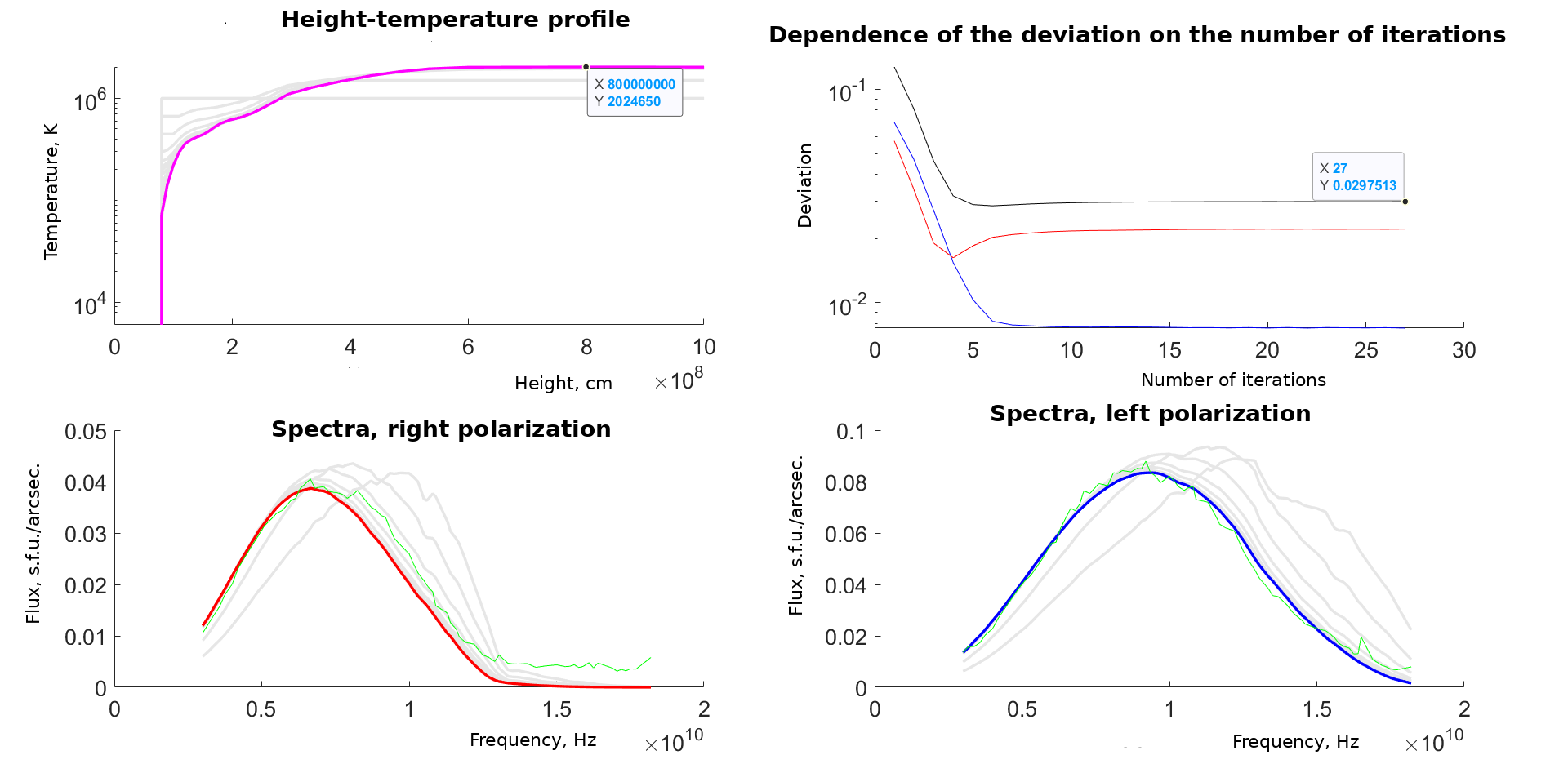}}
\caption{Active region NOAA~11312. Result of reconstruction of the height--temperature profile from the spectrum formed at the maximum of the one-dimensional scan.}
\label{fig:AR11312R1P1}
\end{figure}

Figure~\ref{fig:AR11312R1P1} presents the reconstructed
temperature profile together with the model and observed
microwave spectra in right- and left-circular polarizations.
Good agreement is achieved over the entire frequency range
3--18~GHz; the residual spectral difference does not exceed
2--3\% for each circular polarization separately,
as well as for the total spectrum.

The recovered profile is characterized by values
typical of active regions: coronal temperatures
reach $(2.0$--$2.4)\times10^{6}$~K, and the transition region
is located at heights of approximately $1.5$--$1.8\,\mathrm{Mm}$.
The estimated electron density in the low corona
is of the order of several $10^{9}\ \mathrm{cm^{-3}}$,
consistent with the expected atmospheric parameters
above sunspots.

Testing the dependence of the solution on the choice of the initial
approximation showed that the algorithm converges to similar
height--temperature profiles, indicating the stability
of the obtained solution.

\section{DISCUSSION}

Testing of the algorithm shows that the homogeneous model
provides stable reconstruction of the height--temperature
profile with acceptable accuracy. In the test calculations,
convergence to the prescribed profile is achieved, while for real data
good agreement between the model and observed spectra is obtained.
The highest diagnostic sensitivity is realized
in the 6--15~GHz range, where gyroresonance emission
dominates the formation of the spectrum. Under these conditions,
the method makes it possible to reconstruct the structure
of the upper transition region and lower corona
from microwave spectral observations.

The stability analysis indicates weak dependence of the solution
on the initial approximation: even with substantial shifts
of the initial profile, convergence is preserved.
With added noise up to 3--5\%, the profile is reconstructed reliably;
at higher noise levels, the uncertainty in the position
of the transition region increases, whereas coronal temperatures
remain robustly determined. The parameter $NT$ significantly affects
the spectral shape, especially in the low corona,
and may be considered an additional diagnostic quantity.

At the same time, several limitations of the method should be noted.
The use of a homogeneous model leads to averaging
over the umbra and penumbra structure. The contribution
of the optically thin component ($F^{\mathrm{thin}}$),
particularly at high frequencies, implies that
emission at a fixed frequency is formed
not strictly in the vicinity of a single layer with $\tau_{\nu,p}\sim 1$,
but over an extended height range ($\tau_{\nu,p}<1$).
As a result, the height localization of the emission
is reduced and the sensitivity of the spectrum
to the temperature of individual layers decreases,
which degrades the conditioning of the inverse problem.
Additional uncertainty arises from the treatment of free--free
emission and from possible errors in the magnetic-field reconstruction.

Comparison with previously published results
for the same active region NOAA~11312
demonstrates consistency in the main parameter estimates.
In \cite{Stupishin2018}, coronal temperatures
of 1.5--2.5~MK and densities of order
$10^{9}$--$10^{10}\,\mathrm{cm^{-3}}$
were obtained, in agreement with the present results.
The principal difference lies in the problem formulation.
In the present work, the inversion is formulated
in a strict matrix form and reduced to the solution
of an overdetermined system of linear equations
with regularization. This ensures formal stability
of the solution, allows explicit specification
and variation of the weights of the regularization conditions,
enables quantitative evaluation of the residual and control
of convergence, reduces dependence on the choice
of the initial approximation, and makes the algorithm
reproducible and suitable for systematic application
to different observations.
The proposed scheme also facilitates analysis
of the sensitivity of the solution to noise and discretization parameters,
which was not addressed explicitly in the previous work.
Thus, reconstruction of the height--temperature profile
is transformed from a heuristic iterative procedure
into a formally defined inverse problem
with controlled stability, representing a substantial
methodological development of the approach.

In \citep{Alissandrakis2019}, the same active region was studied
using inversion (DEM) based on
SDO/AIA data. Within the DEM framework, the temperature--density structure
is reconstructed under assumptions of stratification and hydrostatic equilibrium;
the height scale is obtained by integrating the hydrostatic equations
with prescribed boundary conditions. DEM models satisfactorily
reproduce the observed microwave spectra; however, the reconstruction
is limited to the temperature range to which the AIA channels are sensitive
and requires model specification of the lower boundary of the transition region.

In the present work, the height localization of the emitting layers
is determined directly by the gyroresonance formation conditions
through the magnetic-field distribution, providing an independent
diagnostic of the upper transition region and coronal temperature plateau.
Thus, the two approaches rely on different physical assumptions
and employ distinct diagnostic mechanisms.

Comparison of the reconstructed height--temperature profile
with semi-empirical models of the FAL series \citep{Fontenla2009}
shows that the main parameters are consistent
with typical characteristics of active regions:
the position of the transition region and the level of coronal temperatures
fall within the ranges reported in these models.

It should be noted that FAL models are based on inversion
of ultraviolet and X-ray spectral lines formed under NLTE conditions
and often characterized by significant optical thickness.
In such cases, the observed emission is not a direct function
of the local plasma temperature but is determined by the full
radiative transfer problem, including absorption,
reemission, and photon redistribution in frequency
and height. The FAL height--temperature profiles
result from solving the NLTE problem, in which
temperature and density distributions are adjusted
to reproduce the observed spectral lines.

In the microwave range, the emission is continuous
and formed under conditions close to local thermodynamic equilibrium.
The brightness temperature in a layer with optical depth
of order unity is directly related to the electron
temperature of the plasma, allowing more direct
interpretation of observations. Thus,
microwave diagnostics and NLTE spectral-line modeling
represent two fundamentally different and complementary approaches
to reconstructing the thermodynamic structure of the solar atmosphere.

Prospects for further development of the method
are associated with extending the model
beyond one-dimensional averaging. The real structure of a sunspot
is inhomogeneous and includes umbra and penumbra regions
with different magnetic and thermodynamic properties.
The averaged parameterization used here
can be extended to describe these regions separately.
Incorporating multiple scan positions
would allow spatial separation of contributions
from different parts of the sunspot and transition
from one-dimensional averaging to a more realistic description.
A natural next step is application of the method
to two-dimensional brightness-temperature maps
obtained with interferometric radio telescopes,
which would allow abandonment of axial symmetry
and implementation of spatially resolved
inversion of the height--temperature structure.

Another direction of development is a more rigorous treatment
of radiative transfer, including proper description
of the contribution of bremsstrahlung (free--free) emission.
At longer wavelengths, its role increases due to the growth
of the absorption coefficient with wavelength.
At shorter wavelengths, the gyroresonance formation condition
for low harmonics requires higher magnetic-field strengths;
if such values are not reached at the relevant heights,
the relative contribution of the free--free component increases.
Thus, at both ends of the microwave range,
as well as in frequency intervals with weakened gyroresonance,
accurate treatment of bremsstrahlung emission is necessary
for precise reconstruction of the height--temperature profile,
especially in the coronal part.

Finally, comparison of microwave inversion results
with independent diagnostic methods, in particular
DEM analysis based on EUV observations, represents
a promising direction for further research.
Since these approaches rely on different emission mechanisms,
their combined application contributes to a more comprehensive
understanding of the height--temperature structure
of active regions.

\section{CONCLUSION}

In this work, the iterative method for reconstructing the height--temperature profile of the solar atmosphere from multi-frequency microwave observations has been further developed and formally formulated. In contrast to the previously used scheme, the problem is solved as a formal inversion of an overdetermined system of linear equations with regularization, ensuring stability, controllability, and reproducibility of the solution.

Numerical tests with synthetic data demonstrated stable convergence of the algorithm and its behavior in the presence of noise. Application of the method to observations of active region NOAA~11312
(RATAN-600, frequency range 3--18~GHz) made it possible to reconstruct
a height--temperature profile with coronal temperatures
of order $(2.0$--$2.4)\times10^{6}$~K, with good agreement
between model and observed spectra.

Comparison with published results obtained using independent diagnostic approaches
shows agreement in the main characteristics of the temperature structure
in the transition region and lower corona. The adopted homogeneous approximation
defines the limits of interpretation of the reconstructed profile
and indicates directions for further improvement of the method.

\section*{Acknowledgments}

Observations with the SAO RAS telescopes are supported by the Ministry of Science and Higher Education of the Russian Federation. The renovation of telescope equipment is currently provided within the national project ''Science and universities.''

The authors also thank the SDO project team for providing access to the data used in this study.

\section*{FUNDING}
This work was carried out within the framework of the state assignment of SAO RAS approved by the Ministry of Science and Higher Education of the Russian Federation.

\section*{CONFLICT OF INTEREST}
The authors declare no conflict of interest.
\newcommand{\ssr}{Space Science Reviews }

\bibliographystyle{aspb1}
\bibliography{Paper}
\end{document}